\documentclass[10pt,letterpaper,conference]{IEEEtran} 
 
\usepackage[utf8]{inputenc}
\usepackage{graphicx}
\usepackage{amssymb}
\usepackage{amsmath}
\usepackage{booktabs}
\usepackage{subfig}
\usepackage{tabularx}
\usepackage{array}
\usepackage{booktabs}
\usepackage{csquotes}
\usepackage{color}
\usepackage{fixltx2e}

\usepackage[%
  pdftitle={Assessment of Path Reservation in Distributed Real-Time Vehicle Guidance},%
  pdfauthor={Sebastian Senge},%
  pdfsubject={Multi-Agent Systems},%
  pdfproducer={Sebastian Senge},%
  pdfcreator={Sebastian Senge}
  pdfkeywords={vehicle route guidance, distributed system, swarm intelligence}]{hyperref}

\usepackage[noend]{algpseudocode}[2]  
\usepackage{algorithm}

\newcommand{\nai}{na\"{\i}ve }

\begin{document}
    
\title{Assessment of Path Reservation\\ in Distributed Real-Time Vehicle Guidance}

\author{Sebastian Senge\\Department of Computer Science / TU Dortmund, Germany}
\maketitle

\begin{abstract} 
In this paper we assess the impact of path reservation as an additional feature in our distributed real-time vehicle guidance protocol BeeJamA. Through our microscopic simulations we show that \nai reservation of links without any further measurements is only an improvement in case of complete market penetration, otherwise it even reduces the performance of our approach based on real-time link loads. Moreover, we modified the reservation process to incorporate current travel times and show that this improves the results in our simulations when at least 40\% market penetration is possible. %
\end{abstract}
\begin{keywords}
vehicle route guidance, distributed system, swarm intelligence
\end{keywords}

\section{Introduction}\label{sec:Intro}
It is widely acknowledged that traffic congestions in urban road networks have enormous negative impacts with respect to economical as well as ecological aspects~\cite{ExternalCosts}. Hence, there has been extensive research in the field of traffic flow optimization~\cite{Opt} and (on-line) vehicle guidance recently. In \cite{IV12} we proposed a distributed, swarm intelligence-based vehicle guidance approach termed BeeJamA. In contrast to state-of-the-art commercially available systems, based on centralized shortest path algorithms, distributed routing protocols in general have potentially a higher scalability in terms of road segment (link) load update frequency and covered area. The BeeJamA approach borrows from the foraging behavior of honey bees and utilizes a multi-layer hierarichal concept to disseminate routing relevant link load information every second but in a limited vicinity only, similar to the scout-forager behavior of its natural counterpart.  

In \cite{HABC,AntsTITS} two distributed ant-based vehicle guidance approaches have been described. The latter also includes a dynamic path reservation approach, however, it is only evaluated with a complete, or 100\%, (market) penetration. Obviously, this is a factious assumption as different vehicle guidance concepts are and will be used in realistic environments. As the contribution of this paper, we add distributed path reservation to our BeeJamA approach and evaluate the resulting protocol under varying penetration rates on a complex road network with a microscopic traffic simulator.  We will show that a simple and \nai link reservation even decreases the protocol's performance with respect to average travel times if the penetration is not complete. Additionally, we combine the strength of the BeeJamA protocol, fast link load updates, with path reservations and show that this hybrid approach performs better than the \nai one. In our simulations reservations increase the performance of BeeJamA if the penetration is at least 40\%.

The remainder of this paper is structured as follows. In the next section, the basic path reservation concept is described. In Sec.~\ref{sec:BJA}, the plain BeeJamA protocol is briefly explained and the additional reservation procedure is introduced.  Sec.~\ref{sec:Sim} presents the simulation setup as well as the simulation results. Finally, we conclude the findings in Sec.~\ref{sec:Conclusion}.

\section{Path Reservation} \label{sec:PathRes}
Path reservation refers to the registration of vehicles' expected arrival times on the links of their anticipated path.
In subsequent guidance decisions these link reservations can be incorporated to help predicting future link loads. The plain BeeJamA protocol so far intentionally avoids incorporating any kind of traffic load prediction, but tries to offer routing recommendations based on up-to-date traffic information before each intersection in due time. Path reservation, however, requires some temporal lookahead, since the link travel time (the link weight from the perspective of routing protocols) depends on the number of reservations. Hence, one must infer from the number of reservations of a link the expected travel time. The basic tools to express this relation are empirical \emph{density / travel time diagrams} or formal \emph{link performance functions} (LPF). Empirical density / travel time relationships may be learned offline or online by means of machine learning techniques like artificial neural networks. LPFs are equations typically derived from empirical findings as the well known BPR function (\underline{B}ureau of \underline{P}ublic \underline{R}oads, superseded by the Federal Highway Administration of the United States Department of Transportation, cf.~\cite{BPRValues}), given by:

\begin{equation}
t_e(f_e)=t_e^0\left( 1 + \alpha\left(\frac{f_e}{c_e}\right)^\beta \right),
\end{equation}
where $t_e(f_e)$ is the resulting travel time for flow $f_e$ attempting to use link $e$%
, when the free flow travel time is $t_e^0$ and $c_e$ the capacity (in vehicles per hour) the link has been design for. Different values of $\alpha$ and $\beta$ are used to adjust the function to different road types and can be found in~\cite{BPRValues}. 

\begin{figure}[t]
  \centering           
\includegraphics[width=0.35\textwidth]{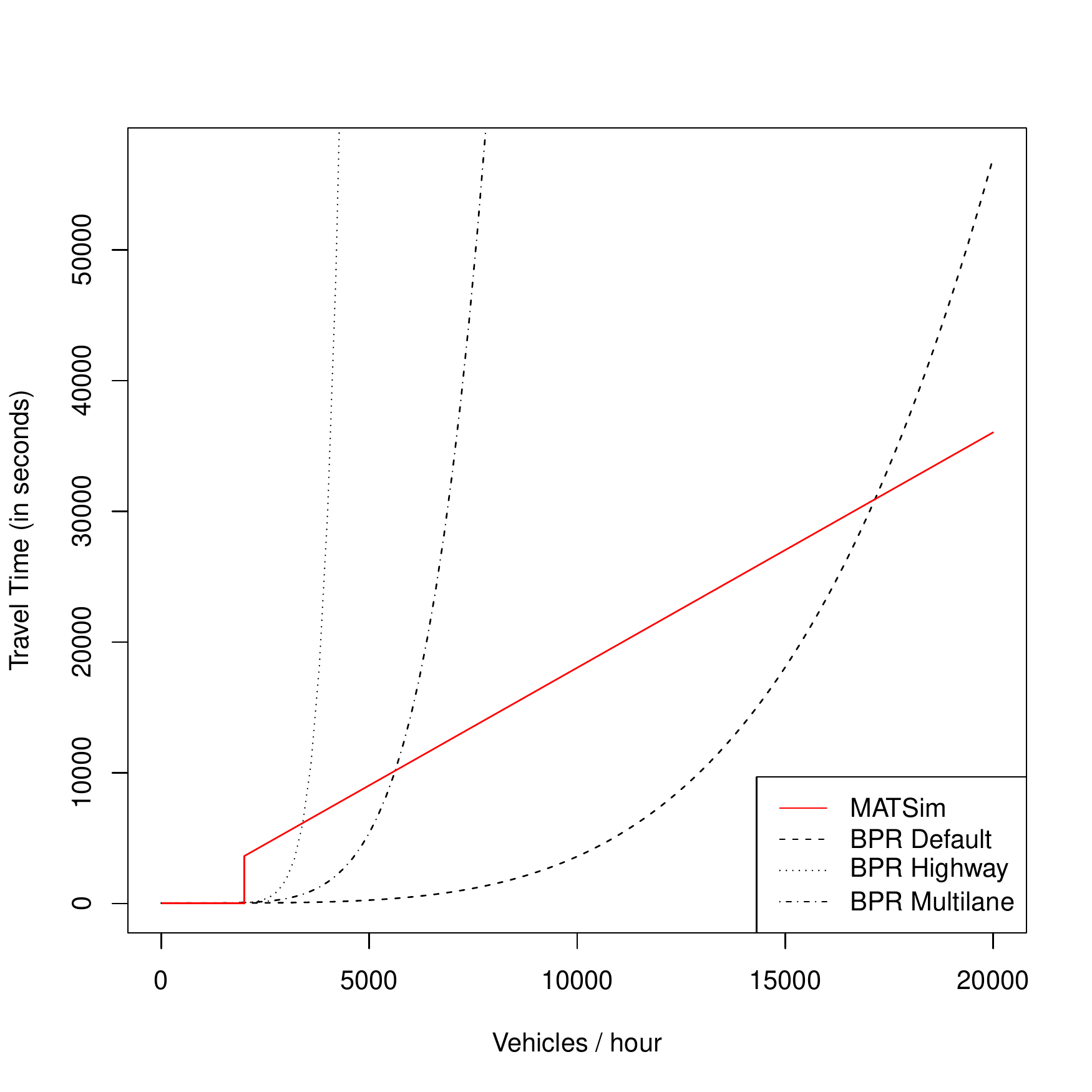} 
  \caption{Link performance function}
  \label{fig:LPF}
\end{figure}

We derived a LPF under the assumption of equidistant headways of the queue-based macroscopic MATSim traffic simulator we used in our simulations from the source code and checked it empirically in a simple test setup where a single link is on the only path from source to destination for all vehicles. The function is given by:
\begin{equation}
t_e(f_e)=
\begin{cases}t_e^0,&f_e < \hat{c}_e\\t_e^0+\left((f_e-1)\frac{3600}{\hat{c}_e}\right), & else,\\
\end{cases}
\end{equation}
where $\hat{c}_e$ is maximum capacity of the link $e$, a value the simulator computes on startup from static properties of a link like type and number of lanes of the road. In general,  the headways must not be not equidistant, thus a pile-up in the incoming queue of a link occurs and the second case ($f_e\geq \hat{c}_e$) describes the travel time by its own. Fig.~\ref{fig:LPF} depicts the MATSim LPF as well as three BPR instances (with $\alpha=0.15,\beta=4$ (default), $\alpha=0.88,\beta=9.8$ (highway), $\alpha=1,\beta=5.4$ (multilane), cf.~\cite{BPRValues} for more details).

Essentially, path reservation then works as follows. With each link a \emph{reservation log} is associated, which consists of \emph{slots} (in this paper with a length of 1 minute). Now suppose a vehicle $v$ is currently at node $n_0$ at time $t_0$ and a path of nodes $(n_0,n_1,\dots, n_m)$ is calculated, where $n_m$ is the vehicle's destination. First, the expected travel time $t_{0,1}$of link ($n_0,n_1)$ is calculated using the link's LPF. Then, the vehicle is \emph{registered} at the link, i.e. each reservation slot for the link in the interval $[t_0, t_{0,1}]$ is incremented and the process is repeated with link $(n_1,n_2)$ until the last link is reached. Thus, subsequent path calculations of other vehicles respect the path choice of $v$, since the former registrations lead to higher LPF values. If a vehicle re-calculates a path during its travel, it has to \emph{update} its previous path reservation, i.e. it removes, advances or postpones the link reservations made so far and/or adds reservations on other links depending on the newly calculated path to reflect changes.

\begin{figure}[t]
  \centering           
\includegraphics[width=0.35\textwidth]{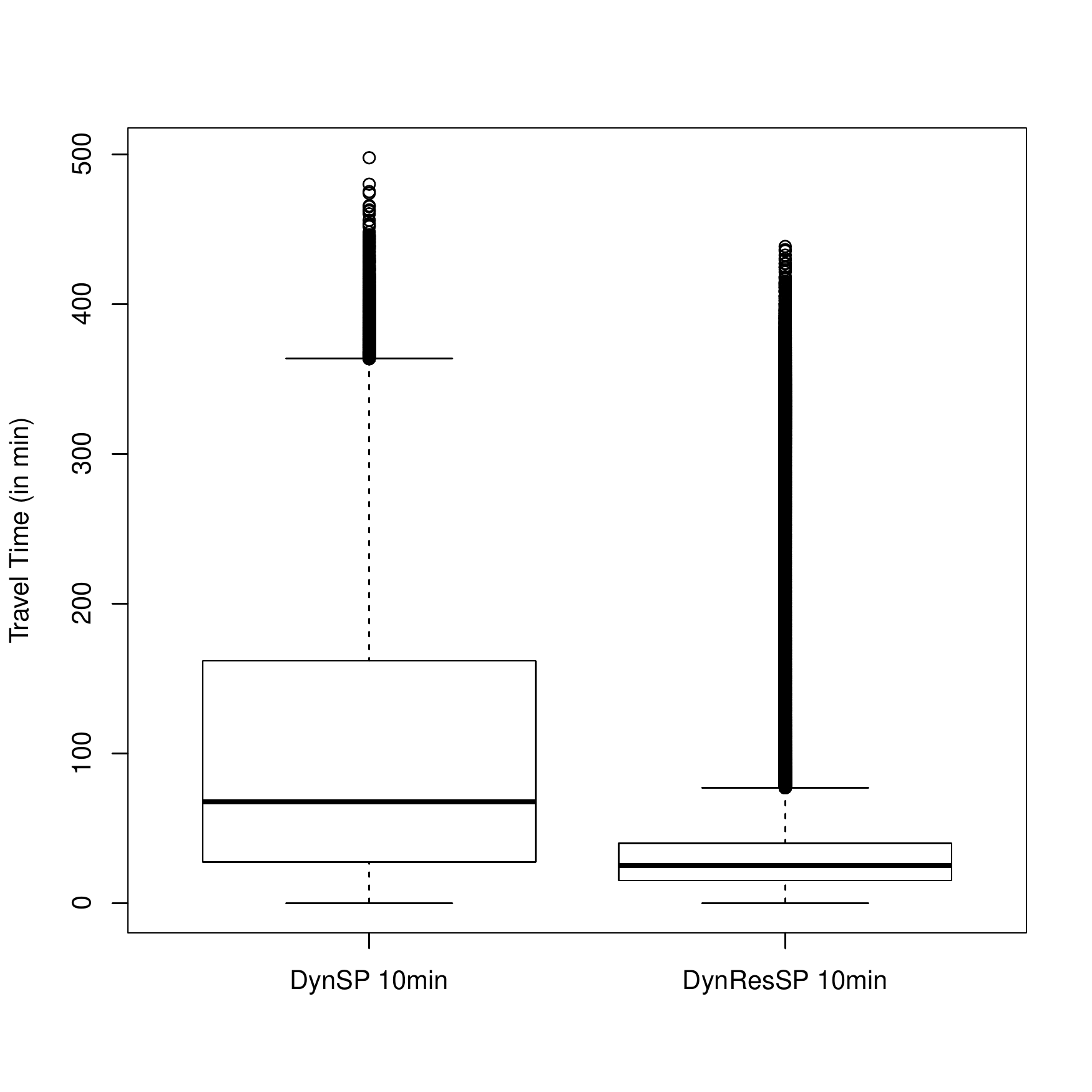}
  \caption{Positive effect  of reservation on travel time distribution}
  \label{fig:Motivation}
\end{figure}

Fig.~\ref{fig:Motivation} may serve as a general motivation for the idea of reserving links on a path. There, we compared a dynamic version of centralized shortest path routing protocol with (DynResSP) and without (DynSP) path reservation, that has access to refreshed averaged travel times over the links of the network every ten minutes. In contrast to the non-reserving version, the DynResSP protocol reserves each time a new path is calculated (in other words each 10min at earliest) the  changed path, i.e. the link is annotated such that it is added when the reserving vehicle will arrive at the beginning of the link and how long it will likely stay on the link. (Detailed elucidations of the simulation setup can be found in Sec.~\ref{sec:Sim}.) The average travel time dropped from about 100min to 42min.  As one can see from the boxplots of the travel time distribution, the lower, medium and upper quartile are considerably better with path reservation, although there are a lot of outliers in case of the path reservation protocol and the maximum values do not differ as desirable. Additionally, the penetration is 100\% and such \emph{perfect} conditions cannot be achieved in realistic environments. However, this simplistic setup makes clear that path reservation may have enormous potential to reduce travel times.

\section{The BeeJamA Protocol}\label{sec:BJA}

The BeeJamA vehicle guidance approach is a distributed routing protocol based on swarm intelligence and is optimized to disseminate least cost path information in large road networks as fast as possible. A comprehensive description of the protocol and additional literature references can be found in~\cite{IV12}. The following sections outline the basic ideas of BeeJamA, which are then extended by path reservation.%

\subsection{Basic Concepts}\label{sec:BasicConcepts}
BeeJamA is a distance vector protocol and continuously floods
  bee agents from each node through the network in the opposite direction of traffic, so-called \emph{upstream scouts}, (see Fig.~\ref{fig:BJAFunc}(a)). Once a node in the network receives a scout, it updates its routing table and floods the agent once again to all predecessors (except the node the bee arrived from). On their travel a scout accumulates the travel time from the links and hence disseminates the cumulated travel time upstream through the network until the hop limit of a scout is reached. A scout's hop limit depends on the hierarchy level of its origin node.  An \emph{hierarchy of overlapping zones} is established by means of a leader election protocol that creates clusters of adjacent nodes and assigns (at least one) \emph{leader} node as cluster representative per cluster. 
The leaders may start a new leader election process among them and thereby create cluster on a higher logical layer level.
The highest layer (with the clusters of largest extent) floods bees with infinite hop limit, lower layers have finite and decreasing hop limits (depending on the configuration which is a tradeoff between accuracy and communication overhead). As a result, it is finally guaranteed that at each node at least one routing table entry leads to at least one cluster representative the destination node is part of. Then, vehicles are always forwarded on the current least cost path to cluster representative of the lowest layer of which scouts have arrived so far at the current node. Finally, the last cluster is the one-element cluster of the destination node itself. 
The protocol is based on a vehicle-to-infrastructure (V2I) architecture, where so-called \emph{navigators} act in lieu of the nodes and links in their limited area of influence. These areas are a node partition of the road network and the navigators maintain routing tables for each of these nodes as well as handle the outgoing and incoming bee agents.

\subsection{Integration of Path Reservation in BeeJamA}

\begin{figure*}
\def\tabularxcolumn#1{m{#1}}
\begin{tabularx}{\linewidth}{XX}
\begin{tabular}{cc}
\quad\qquad\subfloat[]{\includegraphics[width=5.49cm]{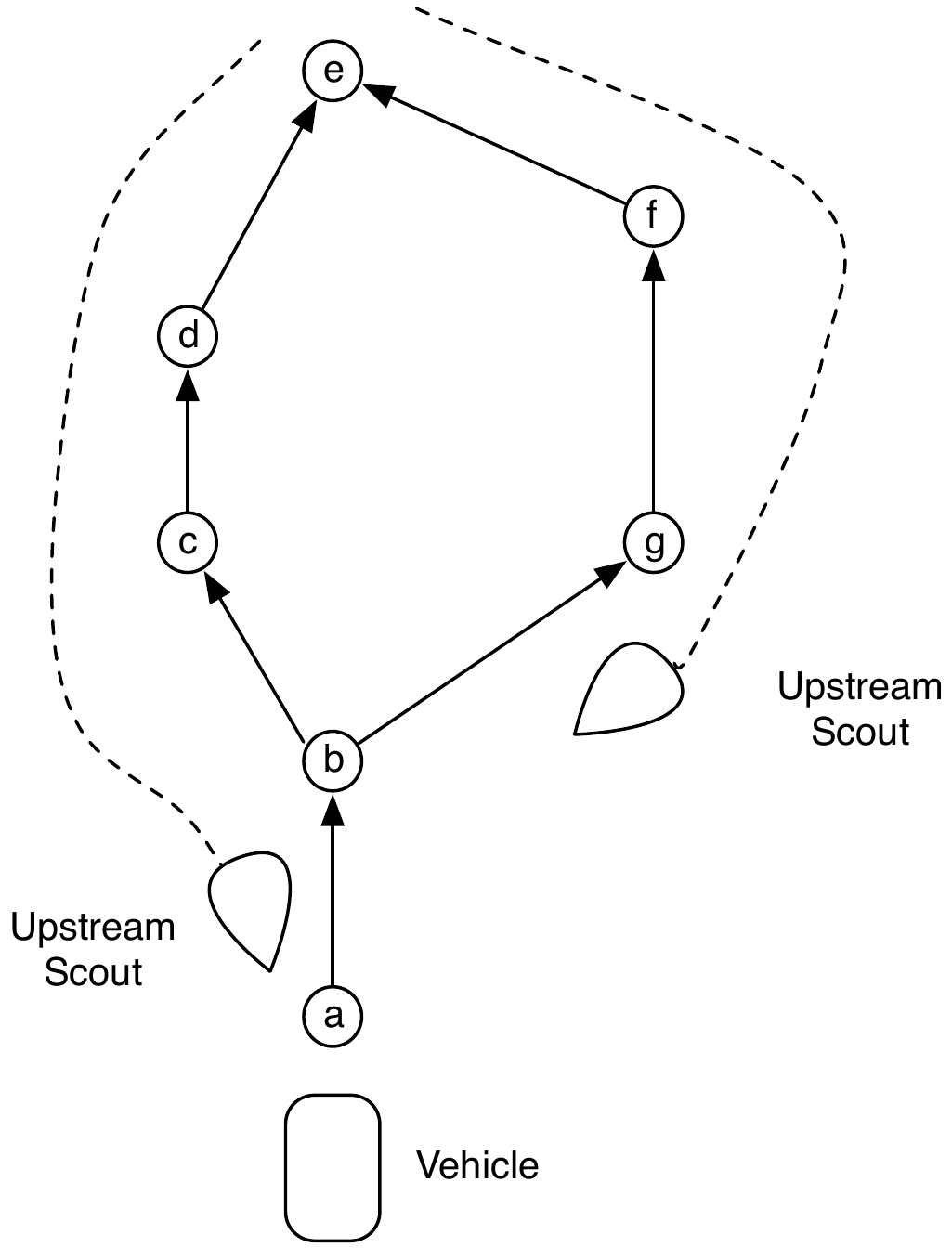}} \qquad\qquad
   & \subfloat[]{\includegraphics[width=6cm]{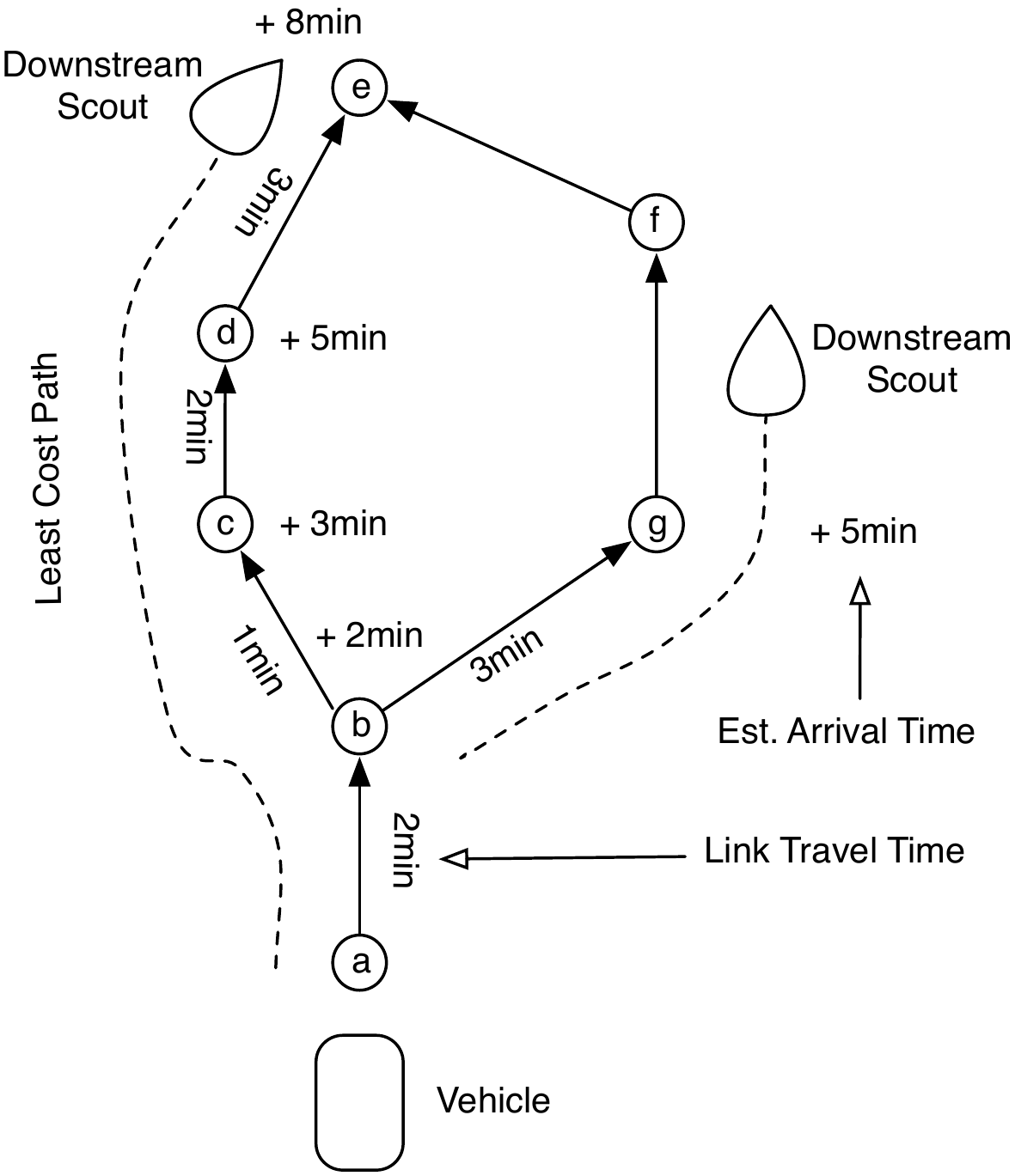}}\\

\quad\qquad\subfloat[]{\includegraphics[height=6.98cm]{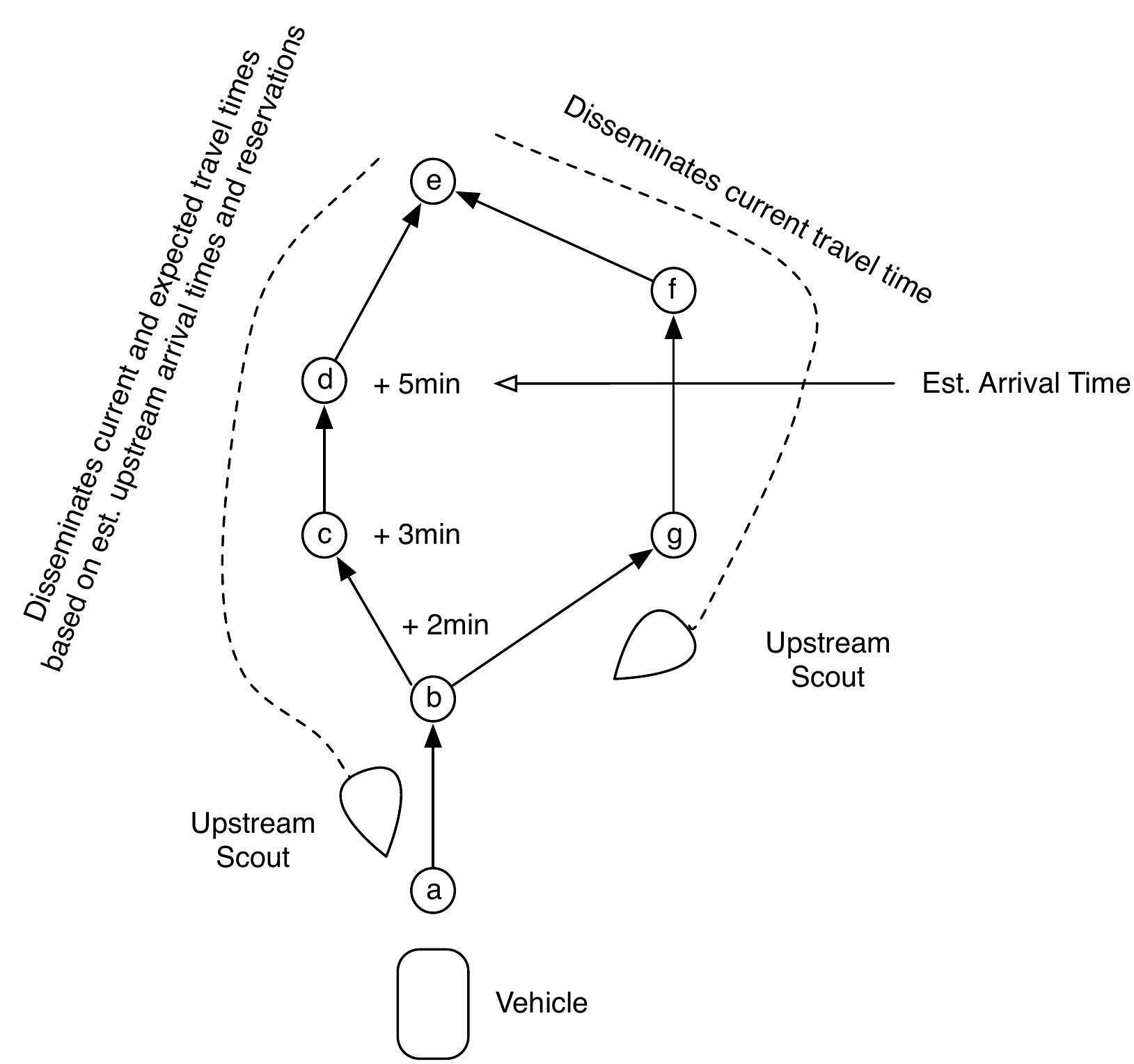}} \qquad\qquad
   & \subfloat[]{\includegraphics[height=6cm]{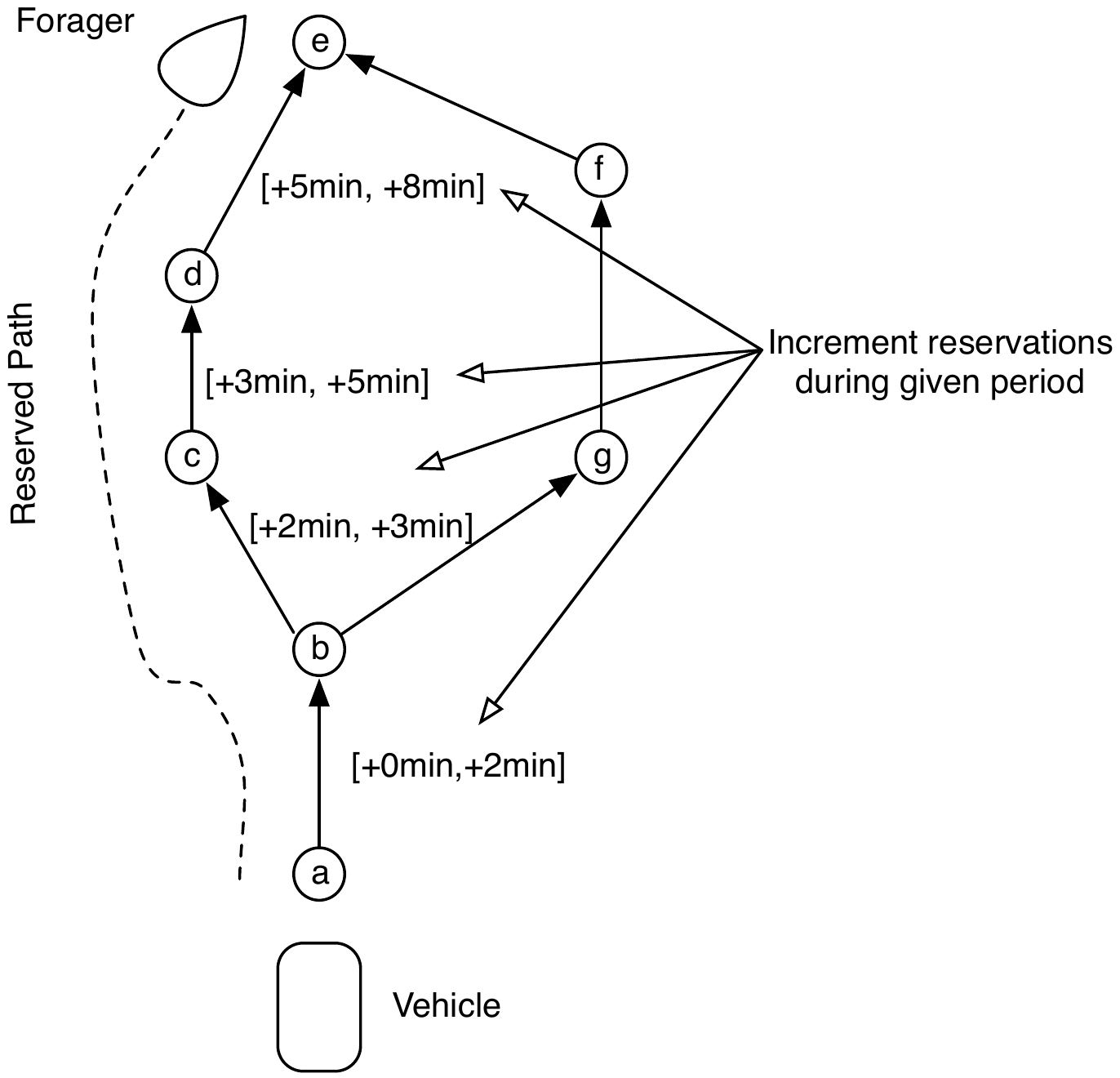}}\\
\end{tabular}

\end{tabularx}

\caption{BeeJamA agents' concepts}\label{fig:BJAFunc}
\end{figure*}

The basic idea to implement path reservation in BeeJamA is that the upstream scouts additionally disseminate the predicted future travel times. Unfortunately, an upstream scout does not know when vehicles (moving downstream) will arrive at the node the upstream scout currently passes. Thus, it is not known for which time in future the LPF should be evaluated. To circumvent this problem, we introduce \emph{downstream scouts} which traverse links in the original direction of traffic flow and work otherwise as their upstream counterparts: cumulating the travel time and updating tables. However, these table entries are not used to forward vehicles but to tell the upstream scouts how long a vehicle currently would need to travel from the downstream scout's origin to the current node. Fig.~\ref{fig:BJAFunc}(b) illustrates this concept. There, node $a$ floods a downstream agent and a situation of the asynchronous process is depicted, where the scout on the left (and least cost) path to $e$ has already reached $e$ (suppose that is the vehicle's destination). Therefore, node $e$ has learned that a vehicle starting now at $a$ would need 8 minutes to arrive at $e$. The next generation of upstream scouts starting at $e$ can now evaluate the LPF for the correct time, i.e. in 5min from now on at node $d$ for link $(d,e)$ (see Fig.~\ref{fig:BJAFunc}(c)). 

The next necessary difference compared to the plain BeeJamA protocol is the use of \emph{foragers}, which are used to increment the number of registered vehicles on a link (see Fig.~\ref{fig:BJAFunc}(d)).  This agent is called a forager, since it announces a consumption of available resources, as a forager bee in reality reduces the amount of available nectar. After the navigator selected the next hop for the requesting vehicle, a forager is emitted on the link towards the selected next hop on the least cost path, i.e. node $b$ in the example. After arriving there, the same navigators computes the next hop for the forager as if it was a vehicle and forwards the agent accordingly. Thus, the forager agent traverses the same path immediately (only delayed by the transmission in the communication network), the slower vehicle would take later if the least cost path does not change in the meantime. When the forager arrives at the destination, the whole path is transmitted back to the vehicle. Subsequent foragers started for this vehicle would possibly result in different paths due to road network load changes, which the vehicle could test for equality and de-register at previously chosen links if the path changed.

By virtue of this approach, a completely distributed and dynamic path reservation protocol emerges. The main drawback of this simple and {\bf n}aive approach (ResBeeJamA-N) is that link travel time is solely derived from the reservations on a link. In case of 100\% penetration this may be and, as the simulation results show, is beneficial in terms of average travel times. For the common situation with incomplete penetration of a single approach, better ways to predict the expected travel time must be used. There exist a lot of approaches for short and long time traffic forecasting~\cite{Learn}
often requiring expensive computation or long learning phases, though. Neither is advantageous in highly dynamic traffic situations, which is why we present a simple hybrid technique of incorporating a LPF and the current mean travel time in the next section.

\subsection{Hybrid Path Reservation}

\begin{figure}[t]
  \centering           
\includegraphics[width=0.35\textwidth]{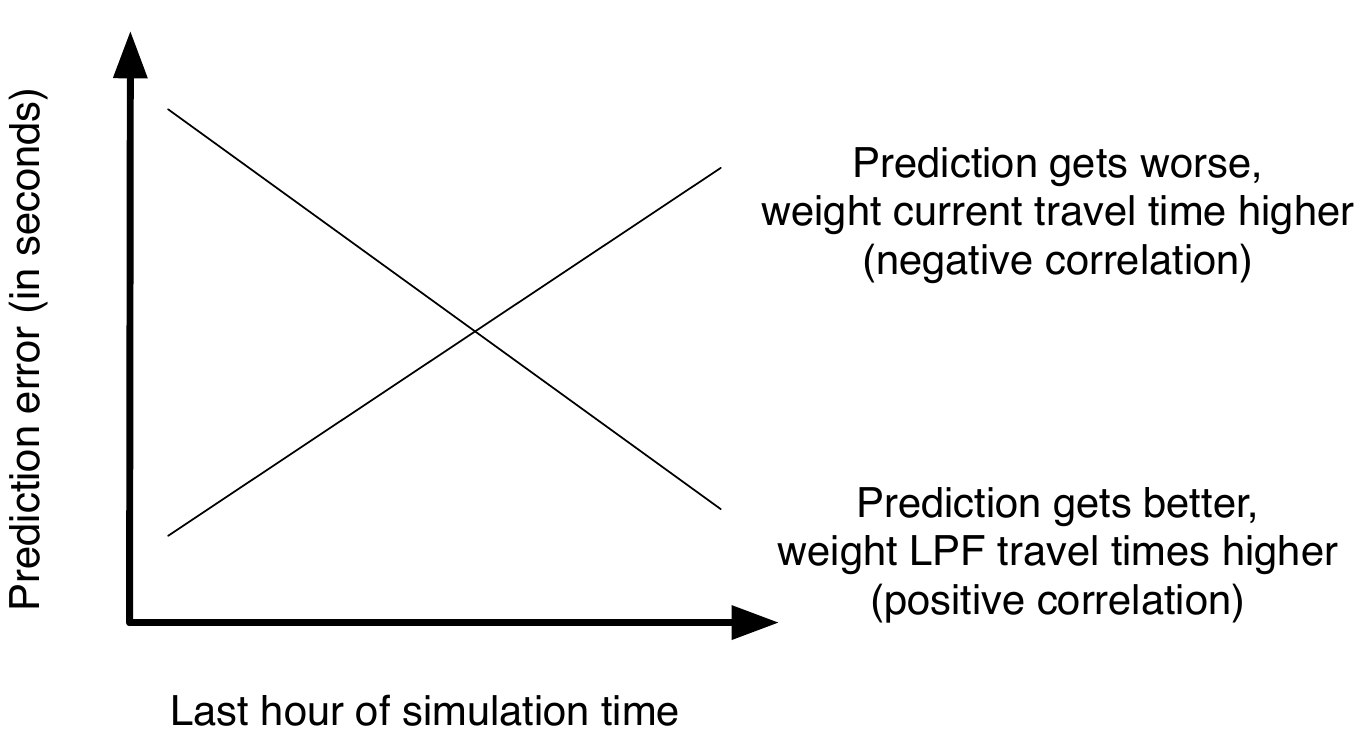}
  \caption{Interpretation of prediction error correlation}
  \label{fig:Cor}
\end{figure}

Our {\bf d}ynamic {\bf h}ybrid approach (ResBeeJamA-DH) evaluates the prediction precision per link and weights the link performance prediction and current load accordingly. The better the prediction over time is, the higher is the impact of the link performance function. The basic notion of this approach is a simple heuristical rule of thumb: if the prediction for a link in the past was precise, the predictions in the future will be precise as well. To assess the past prediction accuracy we compare the difference of past predictions and actual observed travel time on a link. Formally, we apply a regression analysis technique, the Pearson correlation coefficient. Thus, a correlation of time (last hour) and the prediction error (difference of actual and predicted travel time in the last hour) is calculated and serves as a basis for weighting the ratio of the LPF and the current mean travel time.

The \emph{Pearson product-moment correlation coefficient} for random variables $X,Y$ is defined as 
\begin{align}
\rho_{X,Y}=\frac{cov(X,Y)}{\sigma_X\sigma_Y}, \label{eqn:RHO}
\end{align}
where $cov(X,Y)$ denotes the covariance of $X$ and $Y$ and $\sigma_X,\sigma_Y$ the variances of $X$ and $Y$, respectively. Using measures $x_1,\dots, x_n$ and $y_1,\dots, y_n$ the empirical version of Eq.~\ref{eqn:RHO} is:%
\begin{align}
r=\frac{\sum_{i=1}^n(x_i-\overline{x})(y_i-\overline{y})}{\sqrt{\sum_{i=1}^n(x_i-\overline{x})^2} \sqrt{\sum_{i=1}^n(y_i-\overline{y})^2} },
\end{align}
where $\overline{x},\overline{y}$ are the sample means.
During the simulation we take every minute, at time $i$, an observation of the current mean travel time $t_{cur}^{e,i}$ as experienced by the vehicles on each link $e$ and calculate as well a prediction $t_{lpf}^{e,i}$ according the current reservations on $e$. The prediction error $err$ is then simply denoted by the difference $err^{e,i}=t_{cur}^{e,i} - t_{lpf}^{e,i}$. The sample $x_i$ expresses the time of an observation relative to the beginning of the current observation time window (at least one minute, $\{x_1=1\}$ and one hour at a maximum $\{x_1=1,\dots,x_{60}=60\}$).
Obviously, these samples are always strictly increasing, i.e. $x_i<x_{i+1}$. The sample $y_i$ is then the corresponding prediction error. If the prediction error is increasing over time (and there is a linear dependence), a positive correlation $r$ will be found and vice versa. In case of an increasing error on the link, we conclude that the predictions are less reliable and the current travel time may be a better estimation of future travel times (cf. Fig.~\ref{fig:Cor}). %
We express this intuition by 
\begin{align}
t^e=\begin{cases}|r|t_{lpf}^e+(1-|r|)t_{cur}^e, &r\in [-1,0)\\
t_{lpf}, &r=0\\
(1-|r|)t_{lpf}^e+|r|t_{cur}^e, &r\in (0,1]\\
\end{cases}
\end{align}

\section{Simulation studies}\label{sec:Sim}

\begin{figure}[t]
  \centering           
\includegraphics[width=0.4\textwidth]{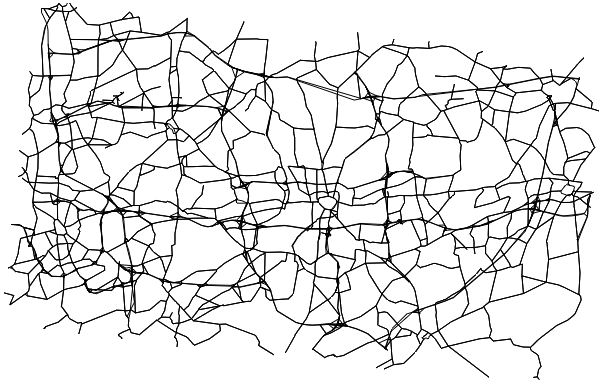} 
  \caption{Ruhr District}
  \label{fig:Graph}
\end{figure}

\begin{figure}[t]
  \centering           
\includegraphics[width=0.35\textwidth]{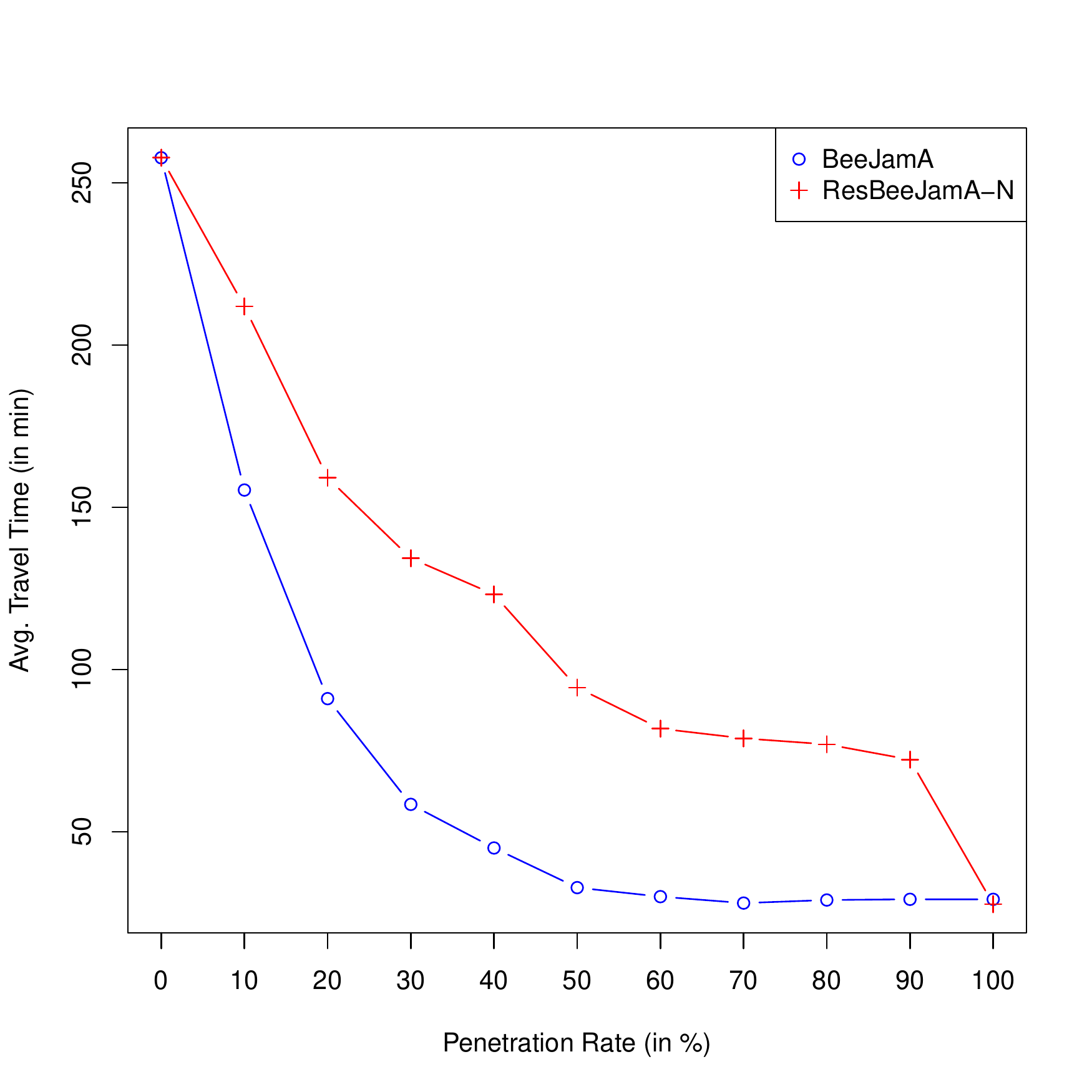}
  \caption{BeeJamA vs. ResBeeJamA-N}
  \label{fig:M1}
\end{figure} 

\begin{figure}[t]
  \centering           
\includegraphics[width=0.35\textwidth]{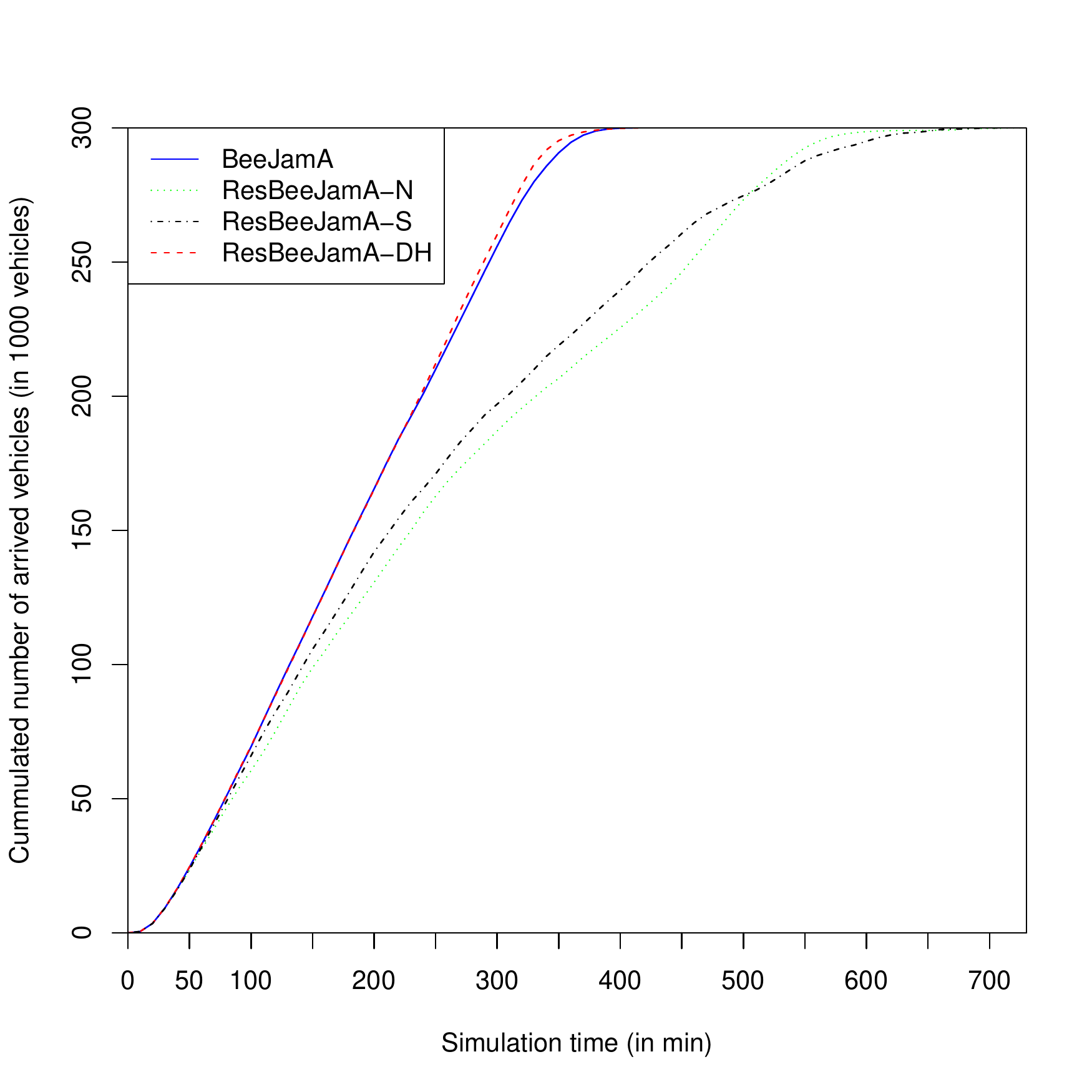}
  \caption{Comparison for a 50\% penetration rate}
  \label{fig:Cum}
\end{figure} 

\begin{figure}[t]
  \centering           
\includegraphics[width=0.35\textwidth]{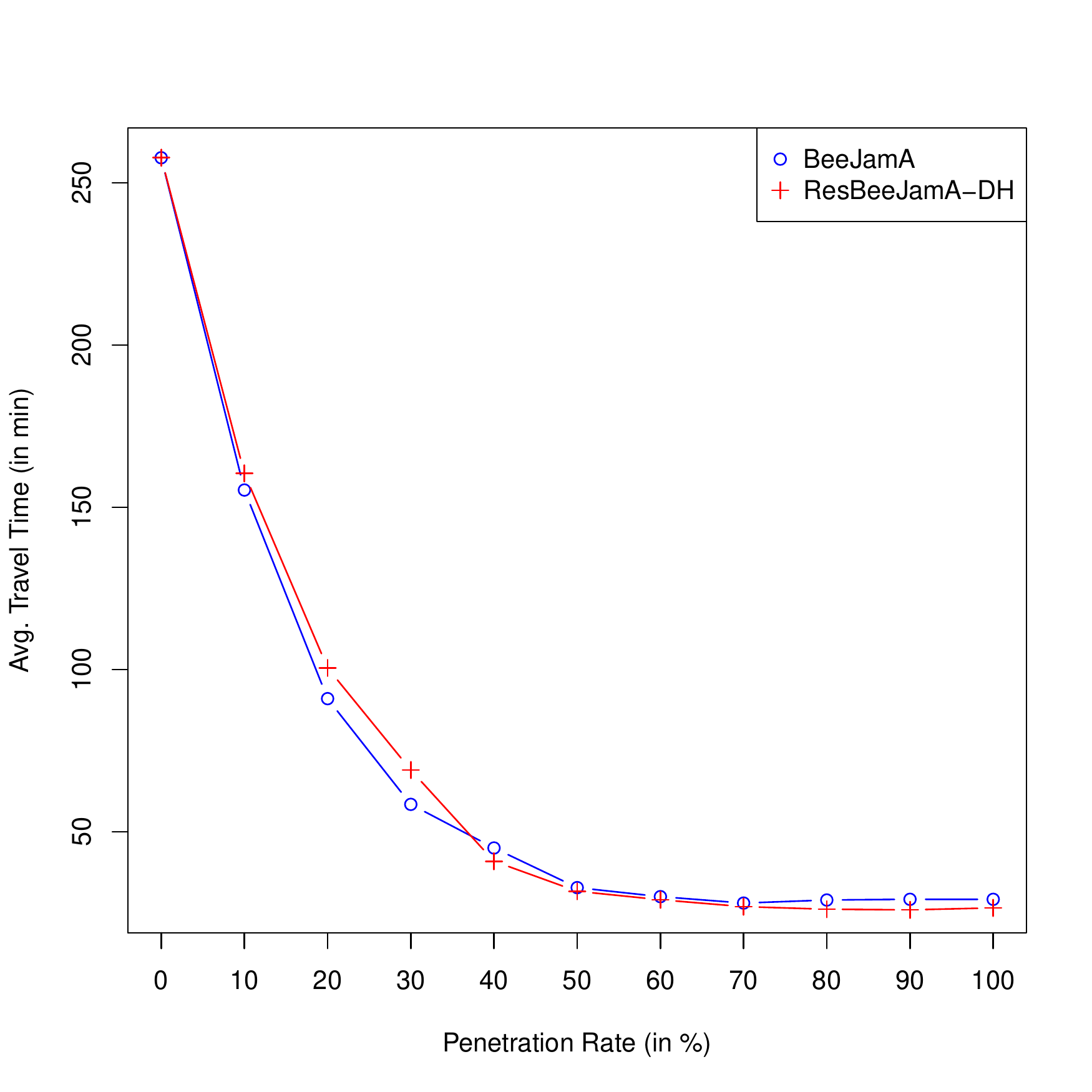} 
  \caption{BeeJamA vs. ResBeeJamA-DH}
  \label{fig:M4}
\end{figure}

\begin{figure}[t]
  \centering           
\includegraphics[width=0.35\textwidth]{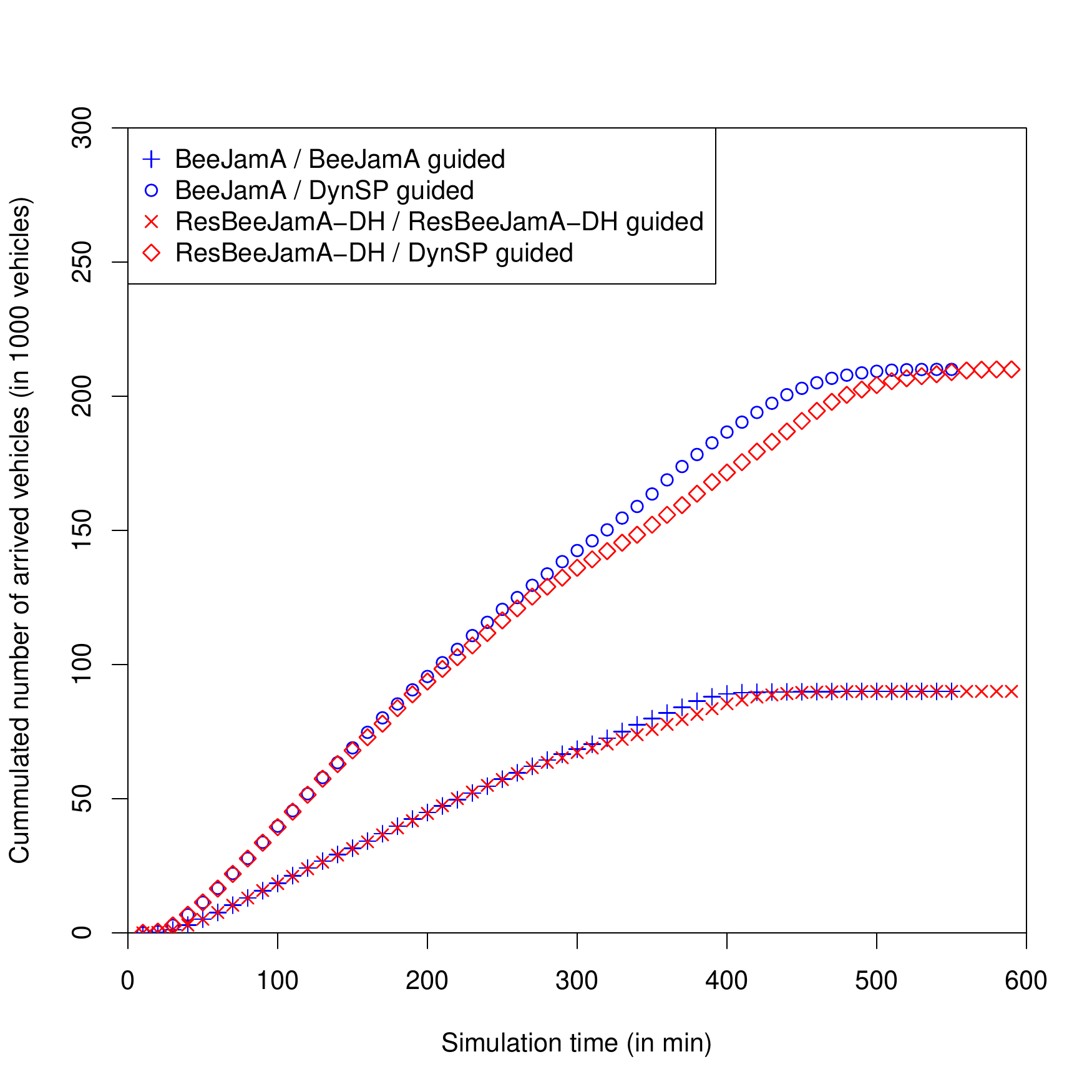} 
  \caption{30\% penetration: BeeJamA vs. ResBeeJamA-DH}
  \label{fig:cum30}
\end{figure}

\begin{figure}[t]
  \centering           
\includegraphics[width=0.35\textwidth]{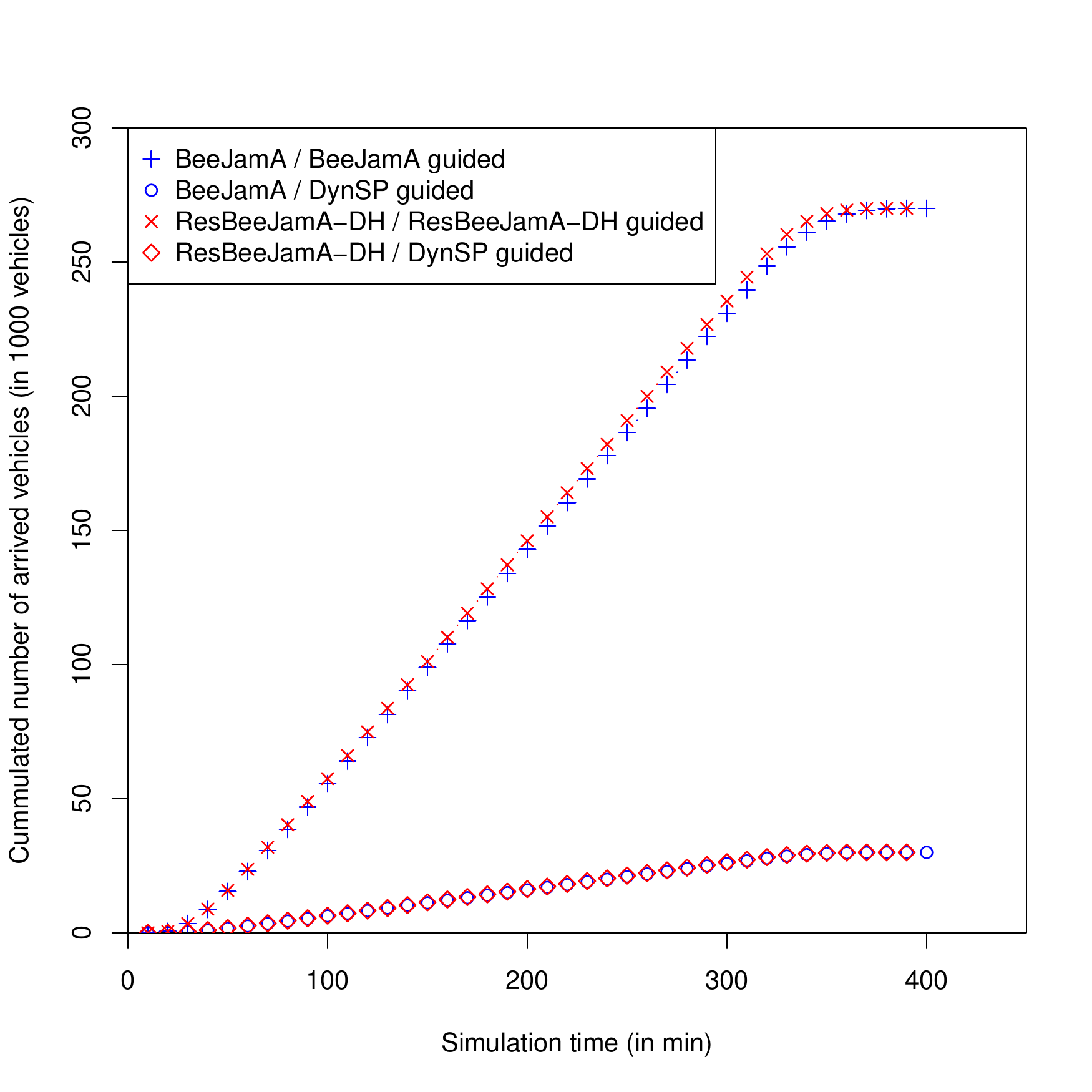} 
  \caption{90\% penetration: BeeJamA vs. ResBeeJamA-DH}
  \label{fig:cum90}
\end{figure}

The guidance protocols has been evaluated using the MATSim microscopic traffic simulator on a road network extracted from Open Street Map. It is a part of the German Ruhr District (see Fig.~\ref{fig:Graph}) with a size of about $40\times 20km^2$. The road network %
exhibits 5616 nodes and 10185 links.
Throughout all simulations an identical setup of the same 300.000 vehicles is reused, where in each of the first five simulation hours 60.000 vehicles start. %

We compared two routing protocols: a dynamic centralized shortest path protocol (DynSP) and our own distributed real-time protocol BeeJamA. The DynSP protocol is powered by a dynamic version of the A* algorithm, i.e. the link weights are updated regularly and the vehicle recalculates the least cost path with the same frequency during its travel. For example if a DynSP 30min protocol is deployed, every 30min the link weights as estimated by the current mean travel time is refreshed. Moreover, all vehicles guided by a DynSP 30min protocol, requests an initial route and afterwards each 30min. In the best case a vehicle recalculates its least cost path right after the link weights were updated, in the worst case the link weights are nearly 30min old and thus maybe outdated.
The reserving modifications of these protocols are denoted as DynResSP and ResBeeJamA and a reserve slot length of 1min is used. The BeeJamA protocol floods each second a generation of scouts, the areas are obtained by a grid partition (with a box side length of 1500m).

In the following simulation studies we will also evaluate the influence of the protocol's \emph{penetration}, i.e. the (market) share of a given protocol. In this paper, remaining vehicles are guided by a DynSP 30min protocol. For example, if ResBeeJamA has a penetration of 30\%, 90.000 vehicles are guided by ResBeeJamA and the remaining 210.000 vehicles by DynSP 30min.

\begin{table}[t]
\renewcommand{\arraystretch}{1.3}
\caption{BeeJamA travel times (in minutes)} 
\label{tbl:ResultsBJA}
\centering 
\begin{tabular}{rccc}
    \toprule
~&~&\multicolumn{2}{c}{ResBeeJamA}\\
Penetration& BeeJamA & N & DH \\
\midrule
0\%&257.74 & 257.74 & 257.74\\
10\%&\fbox{155.32}&211.91&160.46\\
20\%&\fbox{91.02~}&159.10&100.49\\
30\%&\fbox{58.44~}&134.29&69.03\\
40\%&45.00&123.15&\fbox{40.91}\\
50\%&32.78&94.44&\fbox{31.65}\\
60\%&30.00&81.77&\fbox{29.04}\\
70\%&28.02&78.78&\fbox{26.91}\\
80\%&28.96&76.95&\fbox{26.12}\\
90\%&29.18&72.23&\fbox{25.92}\\
100\%&29.17&27.68&\fbox{26.53}\\
\bottomrule
\end{tabular}
\end{table}

In Sec.~\ref{sec:PathRes}, the Fig.~\ref{fig:Motivation} %
was used to motivate the concept of reservation. With the above elucidations, the boxplots belong to the travel time distributions of a DynSP 10min and DynResSP 10min, respectively. As stated in Sec.~\ref{sec:PathRes}, it is obvious that in this example reservation is beneficial. Next, we will evaluate if BeeJamA as a real-time protocol could gain an improvement through reservation. 

Fig.~\ref{fig:M1} depicts the average travel time with an increasing penetration of BeeJamA and ResBeeJamA-N. Obviously, the \nai reservation is not able to improve the performance of BeeJamA, except for the unlikely situation of a complete penetration, where the average travel time dropped from 29.17min to  27.68min (see Tab.~\ref{tbl:ResultsBJA} for details).
Next, in Fig.~\ref{fig:Cum} the cumulative number of arrived vehicles over time under different kinds of BeeJamA for a penetration of 50\% is shown.  Beside ResBeeJamA-N and -DH, we also evaluated a simple static approach (-S), where the weight of the current and LPF travel time is proportional to the penetration rate, i.e. for the 50\% case, the link weight is $2\times t_{lpf}$. The intuitive notion of this weighting is that under a normal distribution assumption, the number of vehicles actually on a link is twice the registered vehicles if the penetration is 50\%.

 The \nai path reservation as well as the static version performs worse, it takes roughly 700min for all vehicles to arrive compared to about 450min for the plain BeeJamA and the ResBeeJamA-DH protocol. Interestingly, the DH reservation is even slightly better and hence we focused in the following simulations on this variant. Fig.~\ref{fig:M4} shows the average travel time over time for the BeeJamA and ResBeeJamA-DH. Even though, both curves are similar, with at least 40\% penetration the path reservation promises benefits for this certain simulation setup. Fig.~\ref{fig:cum30} and Fig.~\ref{fig:cum90} depict the cumulative number of arrived vehicles broken down by (Res)BeeJamA and DynSP protocol for a penetration of 30\% and 90\%, respectively. Once again both protocols behave similar, however, in the case of the lower penetration the plain BeeJamA protocol performs better (resulting in a remarkable difference in the mean travel time of about 8min) and vice versa in the other case (with a difference of about 3min in favour of path reservation).

\section{Conclusion}\label{sec:Conclusion}

We presented a thorough simulative analysis of a distributed, on-line  vehicle guidance protocol with a path reservation extension under varying penetrations. We showed that a \nai implementation, where the reservations are the only source to predict future travel times, does not improve the results in terms of average travel times of our BeeJamA protocol, except in the unlikely case of a 100\% penetration. A hybrid implementation where the current mean travel time over links are incorporated additionally improved the results significantly, though. In future work, will extend the proposed approach to include marginal cost pricing strategies.

\end{document}